\documentstyle[preprint,eqsecnum,aps]{revtex}
\begin{document}
\draft
\preprint{HYUPT-95/2
         \hspace{-28.5mm}\raisebox{2.4ex} {SNUTP 95-032}}
\title{Lie algebra cohomology and group structure of gauge theories}
\author{Hyun Seok Yang and Bum-Hoon Lee}
\address{Department of Physics, Hanyang University, Seoul 133-791, Korea}
\maketitle

\begin{abstract}
\hskip 1.0cm We explicitly construct the adjoint operator of
coboundary operator and obtain the Hodge decomposition theorem
and the Poincar\'e duality for the Lie algebra cohomology of the
infinite-dimensional gauge transformation group. We show that the
adjoint of the coboundary operator can be identified with the BRST
adjoint generator $Q^{\dagger}$ for the Lie algebra cohomology
induced by BRST generator $Q$. We also point out an interesting
duality relation - Poincar\'e duality - with respect to gauge
anomalies and Wess-Zumino-Witten topological terms.
We consider the consistent embedding of the BRST adjoint generator
$Q^{\dagger}$ into the relativistic phase space and identify the
noncovariant symmetry recently discovered in QED with
the BRST adjoint N\"other charge $Q^{\dagger}$.
\end{abstract}


\section{Introduction}
\label{sec:intro}

The theory of gauge fields is based on symmetry principles and the
hypothesis of locality of fields. The principle of local gauge
invariance determines all the forms of the interactions and allows the
geometrical description of the interactions \cite{Utiy56}.
However the quantization of gauge fields leads to difficulties due
to the constraints arising from the gauge symmetry.
These difficulties of the quantization of constrained systems can be
circumvented by the extension of phase space including the anticommuting
ghost variables \cite{Fadd67}.
In this approach, the original gauge symmetry is
transformed into the so-called BRST symmetry in the extended phase
space \cite{BRST,Henn92}. The BRST symmetry will determine all the forms
of the interactions and the algebraic and topological properties
of the fields in the quantum theory \cite{Baul85}.

The question that comes naturally to mind is how we recover the
original gauge invariant space consisting of only physical degrees of
freedom from the extended phase space
with ghosts \cite{Henn92,Baul85,Naka90}
and what is the physical spectrum with the group invariant structure.
In order to study the algebraic and topological structures of
gauge theories, we follow the
point of view of Ref. \cite{Bono83} about the ghost fields and the BRST
transformation. That is, we identify the ghost field with the Cartan-Maurer
form on an infinite-dimensional Lie group $G_{\infty}$ - the group
of gauge transformation - and the BRST generator $Q$
with the coboundary operator $s$ on its Lie algebra ${\cal G}$.
Through these identifications,
we have the natural framework to construct the Lie algebra
cohomology induced by the BRST generator $Q$.
This Lie algebra cohomology will be related to the group invariants
of the configuration space of gauge fields and matter fields.

The organization of this paper is as follows.
In Sec. II, we construct the cochain complex on ${\cal G}$
with values in a ${\cal G}$-module \cite{Cartan,Gold,Choq89}.
With the pairing between
Lie algebra ${\cal G}$ and its dual space ${\cal G}^*$,
we define a chain as an element of the dual space to the cochain
and a dual operation $s_*$ of $s$.
We define a positive-definite inner product and construct an
adjoint operator $s^{\dagger}$ of $s$ using the Hodge duality operation.
We obtain the Hodge decomposition theorem,
Poincar\'{e} duality, and K\"{u}nneth formula analogous
to the de Rham cohomology \cite{Spanier}.
In Sec. III, we show that the adjoint of the coboundary operator can
be identified with the BRST adjoint generator $Q^{\dagger}$ for
the Lie algebra cohomology induced by BRST generator $Q$
and each cohomology class on a polynomial space
is characterized by the gauge invariant polynomials with a particular
group invariant structure imposed on the cochain (or chain) space.
We discuss the physical implications of the Lie algebra cohomology in the
contexts of gauge anomaly and the effective action with the symmetry
group $G$ spontaneously broken to a subgroup $H$.
The Lie algebra cohomology allows us algebraic and topological
characterization of them and provides an interesting
duality relation - Poincar\'{e} duality - between them.
In Sec. IV, we apply this cohomology to QED and QCD.
In order to consider the consistent embedding of the BRST adjoint
generator $Q^{\dagger}$ into the relativistic phase space,
we introduce the nonminimal sector of BRST generator \cite{Henn92}.
Through this procedure, we find the BRST-like N\"{o}ther charge
$Q^{\dagger}$ corresponding to the adjoint of the BRST generator $Q$,
which generates a new kind of noncovariant symmetry in QED
in Refs. \cite{Lave93,Yang1}.
Section V contains discussion and some comments.

\section{Lie algebra cohomology}
\label{sec:coho}

\def\be{\begin{equation}}
\def\ee{\end{equation}}
\def\bea{\begin{eqnarray}}
\def\eea{\end{eqnarray}}
\def\ba{\begin{array}}
\def\ea{\end{array}}

Let $P$ be a principal bundle  with a structure group $G$ (a compact Lie
group with the invariant inner product defined on its Lie algebra
${\it g}$) over a differentiable manifold $M$ (flat Minkowski space or
Euclidean space ${\bf R}^n$).
The gauge transformation group $G_{\infty}$ - an automorphism of $P$-
and its Lie algebra ${\cal G}$ can be identified with
the set of $C^{\infty}$-functions
on $M$ taking values in the structure group $G$ and
its Lie algebra ${\it g}$, respectively.
One defines the dual spaces
${\it g}^*$ of ${\it g}$ and ${\cal G^*}$ of ${\cal G}$
as follows \cite{Choq89}:
\be
<\;x,\;X>=\sum_{a=1}^{dimG} X^a x_a,\;\;
\mbox{for}\; X \in {\it g}\;\;({\cal G}),\;\;
x \in {\it g}^*\;\;({\cal G^*}).\label{digg}
\ee
The spacetime dependence of the elements of $G_{\infty}$, ${\cal G}$,
and ${\cal G}^*$
will be suppressed unless otherwise explicitly indicated and an $L^2$-norm
will be assumed in the inner product (\ref{digg}) between ${\cal G}$ and
${\cal G}^*$ \cite{McMu87}.
Using the pairing between
Lie algebra ${\it g}$ (or ${\cal G}$) and its dual space
${\it g}^*$ (or ${\cal G}^*$), the coadjoint action of $G$ (or $G_{\infty}$)
on ${\it g}^*$ (or ${\cal G^*}$) is defined by
\be
<\;X,\;Ad_g^* x>=<\;Ad_{g^{-1}}X,\;x>
\mbox{for}\; g \in G\;\;(G_{\infty}),\;\; x \in {\it g}^*\;\;({\cal G^*}).
\label{dadg}
\ee

Consider a $p$-cochain $w^p$, an element of $C^p({\cal G};R)$, where
$C^p$ is an antisymmetric $p$-linear map on ${\cal G}$ with values
in a left ${\cal G}$-module $R$
with the ring structure \cite{Cartan,Gold,Choq89,Spanier}.
The space of cochains on ${\cal G}$
is the direct sum of the spaces of $p$-cochains:
\be
C^*=\oplus_{p=0}^{dimG} C^p.
\ee

We introduce on $C^*$ the operators $i(\vartheta)(x)$ and
$\epsilon(\vartheta^*)(x)$ on a point $x\in M$ defined as follows:
\[ i(\vartheta):C^p \rightarrow C^{p-1},\;\;\;\;\;
\forall \vartheta \in {\cal G}\]
by
\be
(i(\vartheta)(x) w^p)(\vartheta_1,\cdots,\vartheta_{p-1})(y)=
 w^p(\vartheta,\vartheta_1,\cdots,\vartheta_{p-1})(y)
\delta(x-y),\;\;\;w^p\in C^p;\label{i}
\ee
and
\[ \epsilon(\vartheta^*) :C^p \rightarrow C^{p+1}, \;\;\;\;\;
\forall \vartheta^* \in {\cal G}^*\]
by
\be
(\epsilon(\vartheta^*)(x)w^p)(\vartheta_1,\cdots,\vartheta_{p+1})(y)=
 \sum_{l=1}^{p+1} (-1)^{l+1} <\;{\vartheta}^*(x),\; {\vartheta}_l(y) >
 w^p(\vartheta_1,\cdots,\hat{\vartheta}_l,\cdots,\vartheta_{p+1})(y),
\label{epsilon}
\ee
where $\;\hat{ }\;$  indicates omission.
Denote by $\{ {\theta}_a \}, a=1,\cdots,N \equiv dimG$,
a basis of ${\cal G}$
and by $\{ {\theta}^{*a} \}$ the basis of ${\cal G}^*$ such that
\be
<\;{\theta}^{*a}(x),\; {\theta}_b(y) >=\delta^a_b\delta(x-y).\label{d}
\ee
Then straightforward calculations using the definitions $(\ref{i})$
and $(\ref{epsilon})$ lead to the following relations \cite{Choq89}
\bea
\ba{l}
\{i(\theta_a),i(\theta_b)\}\equiv i(\theta_a) \circ i(\theta_b)+
   i(\theta_b) \circ i(\theta_a)=0,\\
\{\epsilon(\theta^{*a}),\epsilon(\theta^{*b})\}=0,\\
\{\epsilon(\theta^{*a}),i(\theta_b)\}=
<\;{\theta}^{*a},\; {\theta}_b >{\bf 1}=\delta^a_b{\bf 1},\label{ie}
\ea
\eea
where $\;\circ\;$ denotes the map composition.
Then, for example, the $p$-cochain $w^p \in C^p$ can be constructed using
the operator $\epsilon(\theta^{*})$ as follows
\be
w^{p}=\sum\frac{1}{p!}\underbrace{\epsilon(\theta^{*a}) \circ
\epsilon(\theta^{*b})\circ \cdots \circ \epsilon(\theta^{*c})}
_{p\;\;elements} \phi^{(p)}_{ab\cdots c}, \;\;\;\;\mbox{where}
\;\;\;\phi^{(p)}_{ab\cdots c} \in R. \label{p-cochain}
\ee
It must be kept in mind that the operations in
Eqs. (\ref{i})-(\ref{p-cochain}) must be understood as defined on
a point $x \in M$ and we have omitted delta-function on $M$
in Eq. (\ref{ie}).
This shorthand notation will be used throughout this paper if it
raises no confusion.

Let $s:C^p \rightarrow C^{p+1}$ be
the coboundary operator, i.e., $s^2=0$ \cite{Bono83,Cartan,Gold,Choq89}
defined on $C^*({\cal G};R)$ by
\bea
(sw^{p})(\theta_1,&\cdots&,\theta_{p+1})(x)=
 \sum_{l=1}^{p+1} (-1)^{l+1} \theta_l \cdot
 w^p(\theta_1,\cdots,\hat{\theta}_l,\cdots,\theta_{p+1})(x)\nonumber\\
&+& \sum_{l<n} (-1)^{l+n} w^p([\theta_l,\theta_n],\theta_1,
\cdots,\hat{\theta}_l,\cdots,\hat{\theta}_n,
\cdots,\theta_{p+1})(x),\label{s}
\eea
where a dot means the linear transformation of $R$ defined by an
element of ${\cal G}$. The coboundary operator $s$ can then be expressed
in terms of $\epsilon(\theta^*)$ and $i(\theta)$ as follows
\be
s=\sum_{a=1}^N \int_M\theta_a \cdot \epsilon(\theta^{*a})-
  \sum_{a<b}^N \int\int_M i([\theta_a,\theta_b])
  \circ\epsilon(\theta^{*a})\circ\epsilon(\theta^{*b}),\label{os}
\ee
where the integrations are defined over $M$.

Now we define a chain complex $C$ as the dual space of the cochain
complex $C^*$ using the duality (\ref{digg}) \cite{Gold,Spanier}, namely,
\[ <\;,\;>:C^p\times C_p \rightarrow R\]
by
\be
(w^p,\;v_p)\mapsto <\;w^p,\;v_p>=\int_{v_p}\;w^p, \;\;\;\;\;
w^p\in C^p\;\mbox{and}\;v_p\in C_p,\label{chain}
\ee
where we set $<\;w^p,\;v_q>=0$ if $p\neq q$, and $C^*$ and $C$ are
augmented compleces, that is, $C^p=C_p=0$ for $p<0$ \cite{Cartan,Spanier}.
The duality (\ref{chain}) allows us to define an operator
$s_*:C_p({\cal G}^*;R)\rightarrow C_{p-1}({\cal G}^*;R)$ dual to $s$:
\be
<\;sw^{p-1},\;v_p>=<\;w^{p-1},\;s_* v_p>, \;\;\;\;\;
w^{p-1}\in C^{p-1}\;\mbox{and}\;v_p\in C_p.\label{duals}
\ee
Obviously, Eq. (\ref{duals}) shows us $s^2=0$ implies $s_*^2=0$.
Thus we will identify $s_*$ with the boundary operator acting on
the chains $\{v_p\}$. Of course, the above procedures defining the chain
complex is completely analogous
to the ordinary homology theory \cite{Cartan,Gold,Spanier}.

Let us introduce the Hodge star duality operation whose action on the
cochain space is defined as follows
\be
\ast:C^p \rightarrow C^{N-p}
\ee
by
\be
(*w^p)(\theta_{a_{p+1}},\cdots,\theta_{a_N})=
 \sum \frac{1}{p!} w^p(\theta_{b_1},\cdots,\theta_{b_p})\;
\varepsilon_{\;\;\;\;\;\;\;\;a_{p+1}\cdots a_N}^{b_1 \cdots b_p}.\label{hodd}
\ee
As the de Rham cohomology, we want to define the adjoint
operator $s^{\dagger}$ of $s$ \cite{Gold,Eguchi}
under the new nondegenerate inner product
defined by
\be
(w_1,\;w_2)=\int_{u_N}\;w_1 \wedge *w_2 \label{adjoint}
\ee
with the $N$-chain $u_N$ satisfying $s_* u_N=0$. Then
\be
(sw_1,\;w_2)=(w_1,\;s^{\dagger}w_2),\label{daggers}
\ee
and $s^{\dagger}:C^p \rightarrow C^{p-1}$ is given by
\be
s^{\dagger}=(-1)^{Np+N+1} *\circ s \circ *.\label{dagger}
\ee
For convenience, we have taken the Cartan-Killing metric $g_{ab}$
of the semi-simple Lie subalgebra as positive definite:
\[g_{ab}=-\frac{1}{2}c^{l}_{ad}c^{d}_{bl}=\delta_{ab},\]
where $[\theta_a(x), \theta_b(y)]=c_{ab}^l\theta_l(x) \delta(x-y)$.
The operator $s^{\dagger}$ is nilpotent since $s^{\dagger 2}
\propto * s^2 *=0$. Using the definitions in Eqs. (\ref{dagger}), (\ref{s}),
and (\ref{hodd}), one can determine the action of $s^{\dagger}$ on
a $p$-cochain $w^p$:
\bea
(s^{\dagger}w^{p})(\theta_1,&\cdots&,\theta_{p-1})(x)=-\sum_{l=p}^{N}
\theta_l \cdot w^p(\theta_l,\theta_1,\cdots,\theta_{p-1})(x)\nonumber\\
&-& \sum_{l=1}^{p-1}\sum_{a<b}\;(-1)^{l+1} c_{ab}^l
      w^p(\theta_a,\theta_b,\theta_1,\cdots,
      \hat{\theta}_l,\cdots,\theta_{p-1})(x).\label{sdagger}
\eea
Similarly, the adjoint operator $s^{\dagger}$ can be expressed in
terms of $\epsilon(\theta^*)$ and $i(\theta)$ as follows
\be
s^{\dagger}=-\sum_{a=1}^N \int_M \theta_a \cdot i(\theta_{a})+
  \sum_{a<b}^N \int_M c_{ab}^{\;\;\;c}\; \epsilon(\theta^{*c})
  \circ i(\theta_{a})\circ i(\theta_{b}).
\ee

Let us define an operator $\delta\equiv s\circ s^{\dagger}+
s^{\dagger}\circ s$ corresponding to the Laplacian,
which clearly takes $p$-cochains back into $p$-cochains as
\[ \delta:C^p \rightarrow C^{p}.\]
The straightforward calculation using the Eq. (\ref{ie})
and the Jacobi identity for $c_{ab}^{c}$ leads to the following
expression for the Laplacian $\delta$
\be
\delta=-\int_M (\sum \theta_a \cdot \theta_{a}+
  \sum c_{ab}^{c} \theta_a \cdot \epsilon(\theta^{*c})
\circ i(\theta_{b})+\frac{1}{2}\sum c_{ab}^{c}c_{ae}^{d}
\epsilon(\theta^{*c}) \circ i(\theta_{b})\circ
\epsilon(\theta^{*d})\circ i(\theta_{e})).\label{dlap}
\ee

Considering the formal resemblance to the de Rham cohomology, it will
be sufficient to state, without proof, only the important results
which are necessary for later applications. For mathematical details of
homology and cohomology theory, see Refs. \cite{Cartan,Gold,Spanier}.

We define the $p$-th cohomology group of the Lie algebra ${\cal G}$
by the equivalence class of the $p$-cochains $C^p({\cal G};R)$,
that is, the kernel of $s$ modulo its image:
\be
H^p ({\cal G};R)\equiv Ker^p s/Im^p s, \;\;\; p=0,\cdots,N.\label{cohom}
\ee
Then the nondegenerating inner product (\ref{chain}) provides a natural
pairing between $p$-th cohomology group $H^p ({\cal G};R)$ and
$p$-th homology group $H_p ({\cal G}^*;R)$
\[ H^p ({\cal G};R)\otimes H_p ({\cal G}^*;R) \rightarrow R, \]
so that {\it the inner product (\ref{chain})  establishes the duality of
the vector spaces $H^p ({\cal G};R)$ and $H_p ({\cal G}^*;R)$},
the de Rham theorem \cite{Spanier}.

The following result is the direct consequence of
the positive definiteness of the inner product (\ref{adjoint}):\\
{\it The ``harmonic'' $p$-cochain $w^p\in Harm({\cal G};R)$, i.e.
$\delta w^p=0$ is satisfied if and only if it is exact, i.e. $s w^p=0$
and co-exact, i.e. $s^{\dagger}w^p=0$.}

The adjointness of the operator $s$ and $s^{\dagger}$ under the
nondegenerate inner product (\ref{adjoint}) and their nilpotency lead to
the so-called Hodge decomposition theorem in the cochain space
in a unique way \cite{Gold,Eguchi}:\\
{\it Any $p$-cochain $w^p$ can be uniquely decomposed as a sum of exact,
co-exact, and harmonic forms}, i.e.,
\be
w^p=\delta^p_H\oplus sw^{p-1}\oplus s^{\dagger}w^{p+1},
\;\;\;p=0,\cdots,N,\label{hodc}
\ee
where $\delta^p_H$ is a harmonic $p$-cochain.
The Hodge decomposition theorem (\ref{hodc}) implies {\it the isomorphism
between the $p$-th cohomology space $H^p ({\cal G};R)$ and the $p$-th
harmonic space $Harm^p ({\cal G};R)$}.

The Hodge star operator $*$ maps $C^p \rightarrow C^{N-p}$ and commute
with the Laplacian $\delta$. Thus $*$ induces an isomorphism
\[ Harm^p ({\cal G};R) \approx Harm^{N-p} ({\cal G};R).\]
Consequently, {\it $H^{N-p} ({\cal G};R)$ and $H^p ({\cal G};R)$ are
isomorphic as vector spaces},
\be
H^{N-p} ({\cal G};R) \approx H^p ({\cal G};R). \label{poind}
\ee
This is just the Poincar\'{e} duality \cite{Spanier}.

If the Lie algebra ${\cal G}$ is a direct sum of semi-simple
Lie algebras and/or Abelian $u(1)$ algebras, that is,
${\cal G}={\cal G}_1\oplus {\cal G}_{2}$ and thus each of these algebras
${\cal G}_{\alpha}$ is an ideal of ${\cal G}$, then a total $p$-cochain
$C^p$ will be a sum of a tensor product of cochains
corresponding to each Lie algebra ${\cal G}_{\alpha}$
\[ C^p=\oplus_{q+r=p}\;C_1^q\otimes C_2^r \]
and $w^p\in C^p$ will be given by
\[ w^p=\sum_{q=0}^p w_1^q \times w_2^{p-q},\;\;\; w_1^q\in C_1^q,
\;\;w_2^{p-q}\in C_2^{p-q}.\]
The map $w^p\in C^p$ on ${\cal G}$ is defined by
\[ w^p(\theta_1,\cdots,\theta_{q};\xi_1,\cdots,\xi_{p-q})=
w_1^q(\theta_1,\cdots,\theta_{q})
 w_2^{p-q}(\xi_1,\cdots,\xi_{p-q}),\;\;\;
 \theta_i\in {\cal G}_1,\;\;\xi_i\in {\cal G}_2.\]
Then {\it $H^p ({\cal G};R)$ can be decomposed into a sum of
a product of each $H^q ({\cal G}_{1};R)$ and $H^{p-q}({\cal G}_{2};R)$}:
\be
H^p ({\cal G};R)= \oplus_{q=0}^p [H^q ({\cal G}_1;R)\otimes
H^{p-q}({\cal G}_2;R)].\label{kunn}
\ee
This is known as the K\"{u}nneth formula
for a product space (in our case, a product group
$G_1 \otimes G_2$) \cite{Spanier,Eguchi}.

\section{Group structure of gauge theories}
\label{sec:group}
In this section we will show that the group invariant structure
of constrained system can be described by the Lie algebra cohomology
induced by the BRST generator $Q$ in the algebra of invariant
polynomials on ${\cal G}$ with the generalized Poisson bracket \cite{Henn92},
taking the complete correspondence with the results of Sec. II.
It will provide the algebraic and the topological characterization
with respect to group invariant structures in the gauge theory.

Consider any physical system with gauge transformation group
$G_{\infty}$ and its compact Lie algebra ${\cal G}$ with $N$ generators
$G_a, a=1,\cdots,N$, satisfying the following Lie algebra:
\be
[G_a(x), G_b(y)]=g f^{c}_{ab} G_c(x)\delta(x-y),
\;\;a,b,c=1,\cdots,N. \label{liea1}
\ee
Corresponding to each generator, we introduce a ghost $\eta^a(x)$ and an
antighost $\rho_a(x)$ which satisfy the following
Poisson bracket relations
\be
\{\eta^a, \eta^a\} = \{\rho_a, \rho_b\} = 0,\;\;
\{\eta^a, \rho_b\} = \delta^{a}_{b}.\label{gha1}
\ee
Then we can construct the nilpotent BRST generator \cite{Henn92}
\be
Q=\int_M G_a\eta^a -\frac{1}{2}g\int_M f^{c}_{ab}
\rho_c\eta^a\eta^b,\label{brq1}
\ee
and its nilpotency
\be
Q^2 = 0\label{nilq1}
\ee
follows from the Lie algebra (\ref{liea1}) together with the Jacobi identity.

If one identifies the operators $\epsilon(\theta^{*a})(x)$ and
$i(\theta_a)(x)$ in Sec. II with the ghost $\eta^a(x)$ and
the antighost $\rho_a(x)$ respectively \cite{Choq89},
the expression (\ref{os}) about the coboundary operator $s$ exactly
agrees with the BRST generator $Q$, where structure constants $c_{ab}^l=
g f^{l}_{ab}$ and $G_a$ is any representation for $\theta_a$.
Rewrite the BRST generator as
\be
Q=\int_M (J_a\eta^a -\frac{1}{2}\tau_a\eta^a), \label{brq2}
\ee
where $J_a=G_a+\tau_a$. $\tau_a=g\rho_m f^{m}_{al}\eta^l$ satisfies
the same algebra as $G_a$ and commutes with it.
Then BRST $s$-transformation law with respect to a field ${\cal F}(x)$ is
defined as follows,
\be
s{\cal F}(x)=[Q,{\cal F}(x)\},\label{stran}
\ee
where the symbol $[\;,\;\}$ is the generalized Poisson bracket.
Thus the $s$-transformations with respect to the ghost fields $\eta$
and $\rho$ by $Q$ are
\be
s\eta^a=-\frac{1}{2}g f^{a}_{bc}\eta^b\eta^c,\;\;\;
s\rho_a=J_a. \label{cme}
\ee

According to the Ref. \cite{Bono83}, we identify the ghost field
$\eta(x)$ with a left-invariant Cartan-Maurer form on the group $G_{\infty}$.
With this interpretation of the ghost field $\eta(x)$, the first equation
in Eq. (\ref{cme}) is just the Cartan-Maurer equation with respect to
``exterior derivative'' $s$ for forms $\eta(x)$ on $G_{\infty}$.
It is also obvious that the adjoint operator $s^{\dagger}$ of $s$ introduced
in Sec. II can be constructed in terms of $\eta$ and $\rho$. We define
the corresponding generator by $Q^{\dagger}$ and it is given by
\bea
Q^{\dagger}&=&-\int_M(G^a\rho_a -\frac{1}{2}g f^{ab}_{\;\;\;c}
             \eta^c\rho_a\rho_b),\nonumber\\
           &=&-\int_M(J^a\rho_a -\frac{1}{2}\tau^a\rho_a). \label{cbrq1}
\eea
One can easily check this generator is also nilpotent, i.e.
$Q^{\dagger2}= 0$ as stated in Sec. II.

The generator $Q^{\dagger}$ first appeared in Ref. \cite{Gerv86}
to find the gauge invariant interactions in string theory and then
in Ref. \cite{Holt90} to construct the BRST complex and
the cohomology of compact Lie algebra. The Lie algebra
cohomology in this paper is quite different from
the BRST cohomology constructed
in the paper \cite{Yang2}, so we use the nomenclature,
Lie algebra cohomology, in order to avoid confusion with
the BRST cohomology since these two cohomologies have been often confused
in the literatures.
In fact, the cohomology of Ref. \cite{Holt90}
corresponds to the Lie algebra cohomology in this paper
as long as the spacetime dependences of the Lie group
$G_{\infty}$ and the Lie algebra ${\cal G}$ are fixed.
However, it is necessary to consider the infinite-dimensional
Lie group and Lie algebra in order that
the BRST generator may be viewed as the coboundary operator
for the Lie algebra cohomology \cite{Bono83}.

The $s^{\dagger}$-transformation with respect to a field
${\cal F}(x)$ is defined by
\be
s^{\dagger}{\cal F}(x)=[Q^{\dagger},{\cal F}(x)\}.\label{s^*tran}
\ee
Then the $s^{\dagger}$-transformations with respect to the ghost fields
$\eta$ and $\rho$ are
\be
s^{\dagger}\eta^a=-\;J^a,\;\;\;
s^{\dagger}\rho_a=\frac{1}{2}g f_{a}^{\;\;bc}\rho_b\rho_c.\label{*cme}
\ee
The above equations show that one can identify the antighost $\rho_a$ with
the Cartan-Maurer form with respect to the ``exterior derivative''
$s^{\dagger}$ as well.

Since $Q$ and $Q^{\dagger}$ are nilpotent, it follows that $Q$ and
$Q^{\dagger}$ are invariant by $G_{\infty}$, i.e.
\be
[Q, J_a]=0,\;\;\; [Q^{\dagger}, J_a]=0.\label{brt2}
\ee
One finds that $Q$ and $Q^{\dagger}$ satisfy the supersymmetrylike algebra
that closes into a Laplacian generator $\Delta$
\be
\{Q,Q^{\dagger}\} = -\Delta,\;\; [\Delta, Q]=0,\;\; [\Delta,Q^{\dagger}]=0,
\label{bra}
\ee
where the Laplacian $\Delta$ can be computed in
terms of the Casimir generators \cite{Gerv86}
\be
\Delta=\frac{1}{2}\int_M(J^aJ_a+G^aG_a).\label{lap}
\ee
The operator $\delta:C^p \rightarrow C^{p}$ in Sec. II corresponds
to this generator and it has the exactly same expression as $\Delta$
if it is rewritten in terms of Casimir operators.

Following the same scheme as those in the Refs. \cite{Viol85,Dubo92},
we construct the cochains on ${\cal G}$ spanned by the
polynomial $\omega_{(p)}=Tr\;\eta^p$,
where $\eta=\eta^a T_a$ and $T_a$ is a generator of ${\it g}$.
That is, a p-dimensional cochain $C^p({\cal G};R)$ corresponding to
the Eq. (\ref{p-cochain}) is spanned by elements
of the space of $w^p=\wedge^r\omega_{(p_r)} \cdot \phi \;(\sum p_r=p),$
where $\phi$ is an element of $R$, i.e.
${\cal G}$-module of symmetric polynomials
on ${\cal G}$ without (anti-)ghosts.
Then $\omega_{(p)}=0$ if $p$ is even and $\omega_{(p)}$ is a
``closed'' $p$-form - a $p$-cocycle, i.e.
$s\omega_{(p)}=0$ by Eq. (\ref{cme}).
Notice, for semi-simple groups $G$, $\omega_{(1)}=0$ \cite{Eguchi}.
Let us reexpress the $p$-cochain $w^p$ as the following form:
\be
w^{p}=\sum\frac{1}{p!}\eta^{a_1}\eta^{a_2}\cdots\eta^{a_p}\cdot
\phi^{(p)}_{a_1a_2\cdots a_p}.\label{cochain}
\ee

Note that the results such as Hodge decomposition theorem,
Poincar\'{e} duality, and K\"{u}nneth formula in Sec. II
will be reproduced here in the same manner as well.
In Sec. II, we stated the isomorphism between the $p$-th cohomology space
$H^p ({\cal G};R)$ and the $p$-th harmonic polynomial space
$Harm^p ({\cal G};R)$. Therefore, the BRST invariant polynomial space
can be summarized as the {\it harmonic} polynomial space $\delta w^{p}=0$,
whose solutions are represented by
\be
[G_a, w^p]=0,\label{gsing}
\ee
and
\be
[\tau_a, w^p]=[g\rho_m f^{m}_{al}\eta^l, w^p]=0.\label{ginvs1}
\ee
The second condition reads, in components,
\be
f^{m}_{a[a_1}\phi^{(p)}_{a_2\cdots a_p]m}=0,\label{ginvs2}
\ee
where the square bracket denotes complete antisymmetrization over the
enclosed indices \cite{Holt90}.
The first condition (\ref{gsing}) imposes the $G$-invariance -
$G$-singlet - on the polynomial and the second one imposes very important
constraints about the group invariant structures.
For the $p=0$ and $p=N$, the condition (\ref{ginvs1}) is always satisfied
trivially as long as they are associated with the
$G$-invariant polynomials,
which leads to the conclusion that the zeroth and the $N$-th
cohomology spaces require only the space of $G$-singlet.
For semi-simple groups $G$, there are no solutions
satisfying the condition (\ref{ginvs1}) for $p=1,\;2,\;4$ since there is
no cohomology basis $\wedge^r\omega_{(p_r)}$ to be closed
and for  $p=N-1,\;N-2,\;N-4$ by Poincar\'{e} duality (\ref{poind}),
so that their cohomologies $H^p({\cal G};R)$ vanish.
Note that the gauge group $SU(2)$ is cohomologically trivial so
that the group invariant structure in the $SU(2)$ gauge theory is
similar to eletrodynamics. In this respect, we would like to refer
the interesting analysis \cite{Prokh} which arrives at the same
conclusion under the different approach.
If one $U(1)$ factor is present
(for example, $SU(2) \times U(1),\; U(2)$, etc.),
then $H^1 ({\cal G};R)$ is
non-trivial since $\omega_{(1)}$ is nonzero \cite{Eguchi,Viol85}.
For $G=SU(N),\;N\geq3$, there exist nontrivial cohomologies
$H^3 ({\cal G};R)$ and $H^5 ({\cal G};R)$ whenever
the symmetric polynomials $\phi^{(3)}$ and $\phi^{(5)}$ are proportional
to the structure constants as follows, respectively:
\be
\phi^{(3)}_{abc}=f_{abc}\cdot\phi,\;\;\;\phi^{(5)}_{abcde}=
d_{amn}f_{mbc}f_{nde}\cdot\phi,\label{harm3}
\ee
where $d_{abc}=\frac{1}{2}Tr T_a \{T_b,T_c\}$
and $\phi$ is any $G$-singlet.
These follow directly from the expansion
(\ref{cochain}) \cite{Bono83,Band86} or the Eq. (\ref{ginvs2})
with the Jacobi identity.

It is worth mentioning, for $G=SU(3)$, the nontrivial cohomologies
$H^3 ({\cal G};R)$ and $H^5 ({\cal G};R)$ are related with each other by
Poincar\'{e} duality (\ref{poind}).
The solution of the descent equations corresponding to the Wess-Zumino
consistency conditions in gauge theories \cite{Treiman} shows that
the polynomials $\omega_{(3)}$
and $\omega_{(5)}$ corresponding to the third and the fifth cohomologies
(\ref{harm3}) respectively generate the two dimensional and the four
dimensional gauge anomaly \cite{Bono83,Viol85}
(see also recent analysis \cite{Sore93} by Sorella,
where the cohomology basis $\omega_{(3)}$ and $\omega_{(5)}$
have a fundamental importance on solving the descent equations).
Thus, from the results of these literatures, we can
conclude that 2 and 4 dimensional $SU(3)$ anomalies are related with
each other by the Poincar\'{e} duality; in other words,
the gauge anomaly in two dimensional QCD implies the anomaly in four
dimensional QCD as long as $d$-cohomology
is trivial \cite{Bono83,Viol85,Dubo92}.
This observation is also applied to the problem yielding the general
$G$-invariant effective action \cite{Wein94}
with the symmetry group $G$ spontaneously broken to the
subgroup $H$ since the $G$-invariant effective actions for homogeneous
spaces $G/H$ can be understood as the Lie algebra cohomology problem
of the manifold $G/H$. For example, in the case for $SU(3) \times SU(3)$
spontaneously broken to the subgroup $SU(3)$, the two dimensional
correspondence of the Wess-Zumino-Witten term in four dimensional theory
is the Goldstone-Wilczek topological current \cite{Gold81}.

\section{Cohomology in QED and QCD}
\label{sec:qcd}

In this section, we want to see whether it is possible to find a
corresponding adjoint generator $Q^{\dagger}$ of
the nilpotent N\"{o}ther charge $Q$ in relativistic theories and
what is the role of the adjoint $Q^{\dagger}$ in the Lagrangian
formulation. That is, the solution we want to find out is how to
embed the adjoint $Q^{\dagger}$ of $Q$ into the relativistic phase space.
We showed in Ref. \cite{Yang1} the consistent
nilpotent N\"{o}ther charge $Q^{\dagger}$ exists for Abelian gauge theories
and the generator $Q^{\dagger}$ generates new noncovariant symmetry
and imposes strong constraint on state space.

In order to consider the consistent embedding of the BRST adjoint
generator $Q^{\dagger}$ into the relativistic phase space,
it is necessary to introduce
the nonminimal sector of BRST generator \cite{Henn92,Bata75}.
First, consider the BRST (and anti-BRST) invariant effective QED Lagrangian.
(Our BRST treatments are parallel with those of Baulieu's paper
\cite{Baul85}.)
\bea
{\cal L}_{eff} = &-&\frac{1}{4}F_{\mu\nu}F^{\mu\nu}+ {\bar\psi}
(i\gamma^{\mu}D_{\mu}-m)\psi-\frac{1}{2}
{\bar s}s(A_{\mu}^2+\alpha{\bar c}c) \nonumber\\
= &-&\frac{1}{4}F_{\mu\nu}F^{\mu\nu}+ {\bar\psi}
(i\gamma^{\mu}D_{\mu}-m)\psi + A_{\mu}\partial^{\mu}b+\frac{\alpha}{2}b^2
-\partial_{\mu}{\bar c}\partial^{\mu}c,\label{qedl}
\eea
where $D_{\mu}= \partial_{\mu}+ieA_{\mu}$ is the covarint derivative with
the metric $g_{\mu\nu}=(1,-1,-1,-1).$
The explicit BRST transformations are
\bea
\ba{ll}
sA_{\mu}=\partial_{\mu}c,\;\; & sc=0,\\
s{\bar c}=b,\;\; &  sb=0, \\ \label{qedt}
s\psi=-iec\psi.
\ea
\eea
We introduced an auxiliary field $b$  to achieve off-shell nilpotency of the
BRST (and the anti-BRST) transformation.
Then the nilpotent N\"{o}ther charge generated by the BRST
symmetry reads as
\be
Q = \int d^3x \{(\partial_i F^{io}-J_0)c+b\dot{c}\}, \label{qedq}
\ee
where $J_0$ is a charge density defined by
\be
J_0=e {\bar \psi}\gamma_0 \psi.\label{qedch}
\ee

The constraint functions $G^i$ consist of two commuting groups,
$G^i=(\Phi, b),\;i=1,2$, where $\Phi=\partial_i F^{io}-J_0$
is a Gauss law constraint in
the theory and $b$ is the momentum canonically conjugate to the Lagrange
multiplier $A_0$, so that it generate a gauge
transformation, $\delta A_0$. Thus adding the nonminimal sector in
the BRST generator, the Lie algebra ${\cal G}$ is
composed of a direct sum of two Abelian ideals ${\cal G}_1$ and
${\cal G}_2$ corresponding to the $u(1)$ generators $\Phi$ and $b$,
respectively. In the similar fashion, let the ghost fields split as follows:
\be
\eta^i=(c,\;\; \pi_{{\bar c}}=\dot {c}), \;\;
\rho^i=(\pi_c=-\dot{{\bar c}},\;\; {\bar c}).\label{ghsp1}
\ee
Then the BRST charge $Q$ can be written as
\be
Q=G^i\alpha_{ij}\eta^j,\label{qedbrst}
\ee
where $\alpha_{ij}=\left(\ba{cc}1\;\; 0\\0\;\; 1\ea\right)$.
Since the constraints in the relativistic phase space for Abelian gauge
theories impartially generate $u(1)$ Lie algebras and the K\"{u}nneth
formula (\ref{kunn}) shows $H({\cal G};R)$ is the product of each
$H({\cal G}_{1};R)$ and $H({\cal G}_{2};R)$, we expect it is trivial to
embed the adjoint $Q^{\dagger}$ of the BRST generator $Q$
corresponding to the total Lie algebra ${\cal G}$
including the nonminimal sector into the relativistic phase space.
According to the Eq. (\ref{cbrq1}),
one can guess the form of the generator
$Q^{\dagger}$ must be the following: $Q^{\dagger}=G^i\beta_{ij}\rho^j$.
Note that we have a degree of freedom to the extent of
multiplicative factor in defining the BRST generator $Q$
or the its adjoint generator $Q^{\dagger}$
for a given Lie algebra ${\cal G}$ as long as it does not affect
the nilpotency of $Q$ or $Q^{\dagger}$. Using this degree of freedom
either in the Lie algebra ${\cal G}_{2}$ or in ${\cal G}_{1}$ sector
in defining the adjoint generator $Q^{\dagger}$, we take
the following choices for the matrix $\beta_{ij}$ which will allow the
well-defined canonical mass dimension for $Q^{\dagger}$:
$\beta_{ij}=\left(\ba{cc}1\;\;\;\;\; 0\\0\;\;
-\nabla^2\ea\right)$
or $\left(\ba{cc}\nabla^{-2}\;\; 0\\0\;\;\;\;\; -1\ea\right)$.
These choices make the BRST adjoint $Q^{\dagger}$ the
symmetry generator of the Lagrangian (\ref{qedl}) and
so complete the consistent embedding of $Q^{\dagger}$
into the relativistic phase space.
The former type corresponds to the generator in Ref. \cite{Yang1} and the
latter to the generator in Ref. \cite{Lave93}.

The explicit form of the BRST adjoint generator $Q^{\dagger}$ for the
former type is
\be
Q^{\dagger} = \int d^{3}x \{(\partial_i F^{io}-J_0)\dot{{\bar c}}
+b\nabla^2 {\bar c}\}.\label{qedcbrch}
\ee
Then the explicit transformations defined by (\ref{s^*tran}) are that
\begin{eqnarray}
\ba{ll}
 s^{\dagger}A_{0}=-\nabla^2 {\bar c}, \;\;
 & s^{\dagger}A_i=-\partial_0\partial_{i}{\bar c},\\
 s^{\dagger} c=(\partial_i F^{io}-J_0),\;\;
 & s^{\dagger}{\bar c}=0, \\ \label{qedct}
 s^{\dagger}\psi=ie\dot{\bar c}\psi, \;\;
 & s^{\dagger}b=0.
\ea
\eea
In the Ref. \cite{Yang1}, it has shown that this noncovariant transformation
is a symmetry of the Lagrangian (\ref{qedl}) and that there also exists
the same kind of symmetry in the Landau-Ginzburg and the
Chern-Simons theories.
As discussed in the Ref. \cite{Yang1}, the symmetry generated by
$Q^{\dagger}$ is realized in quite different way compared to the
BRST symmetry: while the gauge-fixing term in the effective QED
Lagrangian (\ref{qedl}), i.e. $A_{\mu}\partial^{\mu}b+
\frac{\alpha}{2}b^2 \rightarrow -\frac{1}{2\alpha}(\partial_{\mu}A^{\mu})^2$,
remains invariant under the transformation (\ref{qedct}),
the variation from the ghost term is canceled up to the total
derivative by the variation
from the original gauge-invariant classical Lagrangian which
remains invariant under the BRST transformation (\ref{qedt}).
These differences in the way of realizing the symmetries imply
that the BRST adjoint
symmetry can give the different superselection sector
from the BRST symmetry \cite{Lave95} (as it is also seen from
the Hodge decomposition theorem (\ref{hodc})
which is a canonical decomposition into a direct sum of linearly
independent subspaces) unlike the recent comment \cite{Rive95}.

If we choose, instead, the matrix $\beta_{ij}=\left(\ba{cc}
\frac{1}{\nabla^2}\;\; 0\\ 0\;\; -1\ea\right)$ in the Eq. (\ref{qedcbrch}),
we will obtain the nonlocal symmetry in Ref. \cite{Lave93}. Of course,
in this case, we must impose the good boundary conditions on fields.
But there is no reason to introduce the nonlocality and it seems
unnatural since the generator $Q^{\dagger}$ must be the adjoint of
the generator $Q$ of the {\it local} gauge transformation.

The adjoint generator in the configuration space can be understood as the
generator of transformation consistent with the gauge fixing condition
\cite{Lave93,Yang1}. Thus, in the configuration space,
there may not exist the global expression of the adjoint generator
$Q^{\dagger}$ of non-Abelian gauge theory compatible with the gauge
fixing condition on account of the topological obstructions
such as Gribov ambiguity \cite{Grib78}.
But it does not imply that there can not exist the local expression
of $Q^{\dagger}$, because the difficulty posed by
the Gribov ambiguity can be avoided \cite{Singer}
by finding a local cross section on a finite local
covering and using the Faddeev-Popov trick locally.
Nevertheless, it seems a nontrivial problem to find the solution
for the consistent embedding into the relativistic phase space
for the non-Abelian gauge theory such as QCD.
This problem remains to be future work.
We want to focus our attention about the
construction of $su(3)$ Lie algebra cohomology in QCD.

Consider the BRST (and anti-BRST) invariant effective QCD Lagrangian:
\bea
{\cal L}_{eff} = &-&\frac{1}{4}F^{a}_{\mu\nu}F^{a\mu\nu}+ {\bar\Psi}
(i\gamma^{\mu}D_{\mu}-M)\Psi-\frac{1}{2}
{\bar s}s(A^a_{\mu}A^{a\mu}+\alpha{\bar C}^aC^a) \nonumber\\
= &-&\frac{1}{4}F^2_{\mu\nu}+ {\bar\Psi}
(i\gamma^{\mu}D_{\mu}-M)\Psi
+ A_{\mu}\partial^{\mu}B+\frac{\alpha}{2}B^2+
\frac{\alpha}{2}gB[C,{\bar C}]\nonumber\\
&-&\partial_{\mu}{\bar C}D^{\mu}C+\frac{\alpha}{2}g^2[{\bar C},C]^2,
\label{qcdl}
\eea
where quark fields $\Psi$ are taken to transform according to the
fundamental $SU(3)$ representation,
the Yang-Mills vector potential $A_{\mu}$, a pair of anticommuting
ghosts $C$, ${\bar C}$ and
the auxiliary field $B$ take values in the adjoint
representation of a $SU(3)$ Lie group. The QCD Lagrangian (\ref{qcdl})
is invariant with respect to the following
BRST transformations \cite{Baul85}:
\begin{eqnarray}
\ba{ll}
 sA_{\mu}=D_{\mu}C,\;\;\; &  sC=-\frac{g}{2} [C,C],\\
 s{\bar C}=B, \;\;\; & sB=0,\\  \label{qcdt}
 s\Psi=-gC\Psi.
\ea
\eea
$D_{\mu}$ defines the covariant derivatives of $SU(3)$ Yang-Mills
symmetry group.
The corresponding conserved nilpotent BRST generator is
given by
\be
Q = \int d^3x \{(D_i F^{io}-J_0+g[\dot{\bar C},C])^aC^a
+B^a(D_0C)^a -\frac{1}{2}g [\dot{\bar C},C]^aC^a\}, \label{qcdq1}
\ee
where $J_0^a$ is a matter color charge density defind by
\be
J_0^a=-ig {\bar \Psi}\gamma_0 T^a \Psi.\label{qcdcch}
\ee

The constraint functions $G^A$ are composed of two commuting groups,
$G^A=(\Phi^a, B^a)$, where $\Phi^a=(D_i F^{io}-J_0)^a$  is the original
Gauss-law constraints
in the theory generating $su(3)$ Lie algebra:
\be
[\Phi_{a}, \Phi_{b}]=g f^{c}_{ab} \Phi_{c},\label{qcdlie}
\ee
and $B^a$s are the momenta
canonically conjugate to the Lagrange multipliers $A^a_0$ and generate
$u(1)$ Lie algebras. In the similar fashion as QED, one can split the ghosts
as follows:
\be
\eta^A=(C^a,\;\; \Pi^a_{{\bar C}}=(D_0 C)^a), \;\;
\rho^A=(\Pi^a_C=-\dot{\bar C^a}, \;\;{\bar C}^a).
\label{qcdghsp}
\ee
Note that $s\Pi^a_{{\bar C}}=0$, so that we can identify the ghost
$\Pi^a_{{\bar C}}$ with the Cartan-Maurer form on $U(1)$ group.
Of course, the BRST generator $Q$ in Eq. (\ref{qcdq1}) is exactly
same form of the Eq. (\ref{brq1}).
Let us rewrite the BRST generator $Q$ as the form of
the Eq. (\ref{brq2})
\be
Q = \int d^3x \{J_aC^a+B_a\Pi^a_{{\bar C}}
-\frac{1}{2}\tau_a C^a\},\label{qcdq2}
\ee
where the generator $J^a$ and the generator of the ghost
representation $\tau_a$ \cite{Gerv86} are given by
\be
J^{a}=(D_i F^{io}-J_0+g[\dot{\bar C},C])^a=\Phi^a+\tau^a,\;\;\;
\tau^a=g[\dot{\bar C},C]^a.\label{qcdj}
\ee
The generators $J_a$ and $\tau_a$ satisfy the same $su(3)$ algebra:
\be
[J_{a}, J_{b}]=g f^{c}_{ab} J_{c},\;\;\;
[\tau_{a}, \tau_{b}]=g f^{c}_{ab} \tau_{c}. \label{qcdtau}
\ee

Since the two groups of the constraint functions $G^A=(\Phi^a, B^a)$ commute
with each other, the total Lie algebra ${\cal G}$
including the nonminimal sectors $B^a$ is composed of the $su(3)$
non-Abelian ideal and the eight $u(1)$ Abelian ideals:
\be
{\cal G}=\oplus su(3)\oplus^{8}_{\alpha=1} u(1)_{\alpha}. \label{qcdg}
\ee
In order to construct only the cohomology of
the color $su(3)$ Lie algebra for the reason explained above,
we drop the Abelian sectors from the BRST generator $Q$ through the
direct restriction on the cochain space (\ref{cochain}), in other words,
considering only $su(3)$ sub-cochain complex.
The BRST adjoint $Q^{\dagger}$ defined on the cochain
$C^*(su(3);R)$ is equal to
\be
Q^{\dagger}=-\int d^3x \{J^a\Pi^a_C -\frac{1}{2}\tau^a \Pi^a_C \}.
\label{qcdcq}
\ee
Then the Laplacian $\Delta$ of the $su(3)$ subalgebra sector
can be represented in terms of the generators $J_a$
and the original constraints $\Phi_a$
\be
\Delta=\frac{1}{2} \int d^3x \{J^aJ_a+ \Phi^a\Phi_a\}, \label{qcdlap}
\ee
which is equal to the expression given by Eq. (\ref{lap})
for $su(3)$ cohomology.
Thus the harmonic polynomials of the $su(3)$ algebra sector
must satisfy the following conditions,
\be
[\Phi^a, w^p]= [(D_i F^{io}-J_0)^a, w^p]=0,\;\;
   a=1,\cdots,8,\label{qcdgsing}
\ee
and
\be
[\tau^a, w^p]=[gf^a_{bc}\dot{\bar C^b}C^c, w^p]=0,\;\;
 a=1,\cdots,8.\label{qcdginvs}
\ee
{}From the arguments in Sec. III, we see that the solutions of
Eqs. (\ref{qcdgsing}) and (\ref{qcdginvs}) exist
trivially for $p$=0 and $p$=8 as long as they are given by
the gauge invariant polynomials because they are singlets
under the adjoint representation of the $su(3)$ Lie algebra.
But the cohomologies $H^p(su(3);R)$ for $p=1,\;2,\;4,\;6$, and $7$ vanish.
For $p$=3 and 5, there always exist non-trivial cohomologies
$H^3(su(3);R)$ and $H^5(su(3);R)$ whose structures
are given by Eq. (\ref{harm3}) and
they are related with each other
by the Poincar\'{e} duality (\ref{poind}).
Since the Lie algebra cohomology proves the nontrivial property
of group invariant structures, the nonvanishing Lie algebra
cohomologies $H^p(su(3);R)$ can be related to the
gauge invariants in $SU(3)$ gauge theory.
It remains to investigate the deep relation between the gauge
invariant configuration of gauge and matter fields in the
spacetime and the Lie algebra cohomology.

\section{Discussion}
\label{sec:disc}

We have constructed the Lie algebra cohomology of the group of gauge
transformation and obtained the Hodge decomposition theorem
and the Poincar\'{e} duality.
As long as a Lie algebra has a nondegenerate Cartan-Killing
metric so that the underlying manifold is orientable,
we can always define a unique (up to a multiplicative factor)
adjoint of the coboundary operator
under a nondegenerate inner product using a Hodge duality.
However, for Lie algebras such as the Virasoro algebra
for which no Cartan-Killing metric exists, the adjoint can not
be unique. Indeed, for the Virasoro algebra, the adjoint of BRST generator
defined by Niemi \cite{Niem} is different from ours and
that in Ref. \cite{Gerv86}.

We also considered the consistent extension of
the Lie algebra cohomology into the relativistic phase space
in order to obtain the Lagrangian formulation.
In order to do that, we extended the Lie algebra
by including the nonminimal sector of BRST generator.
The adjoint $Q^{\dagger}$ constructed through this
procedure generates the noncovariant local or nonlocal symmetry
in QED in Refs. \cite{Lave93,Yang1}.
We have pointed that there is no reason to introduce
the nonlocality necessarily
and it seems unnatural since the generator $Q^{\dagger}$ must be
the adjoint of the BRST generator $Q$ generating local gauge transformation.
But, in the configuration space, the adjoint $Q^{\dagger}$
compatible with the gauge fixing condition can not exist globally
for the non-Abelian gauge theory due to the topological obstructions
such as Gribov ambiguity. As explained in Sec. IV,
the adjoint $Q^{\dagger}$ in the non-Abelian gauge theory
can exist locally (or perturbatively), so that it can generate
new symmetry at least locally (or perturbatively).
So it will be interesting to study the role
of the symmetry transformation generated by the generator $Q^{\dagger}$
and the Ward identity of this symmetry
in the local (or perturbative) sense.

Note that the Lie algebra cohomology constructed here
is quite different from the BRST cohomology
in Refs. \cite{Naka90,Yang2,Spie87}.
In the two cohomologies, the role of ghost fields is
quite different and each inner product to obtain Hodge theory is defined
by the definitely different schemes. It can be shown \cite{Yang3} that
there is no paired singlet in the BRST cohomology
so that higher cohomologies
with nonzero ghost number vanish as long as the asymptotic completeness is
assumed. Therefore the ghost number characterizing cohomology classes in this
paper has different meaning from the ghost number of state space.
The distinction between the BRST cohomology and the Lie algebra
cohomology will be further clarified \cite{Yang3}.

In QCD, there are nontrivial cohomologies $H^p(su(3);R)$ for $p=0,\;8$
and $p=3,\;5$ and they are, respectively,
related to each other by the Poincar\'{e} duality. Since the Lie
algebra cohomology proves the nontrivial property of group invariant
structures, the nonvanishing Lie algebra cohomologies $H^p(su(3);R)$
may be deeply related to the colorless combination
of $SU(3)$ color charges which satisfy the $su(3)$ Lie algebra.
Then it will be very interesting to investigate the relation between
the color confinement and the $su(3)$ cohomology.

\section*{ACKNOWLEDGEMENTS}
This work was supported by the Korean Science and Engeneering Foundation
(94-1400-04-01-3 and Center for Theoretical Physics) and
by the Korean Ministry of Education (BSRI-94-2441).

\end{document}